\documentclass{pasj01}
\usepackage{url}

\begin{document}
\SetRunningHead{Tsujimoto et al.}{X-ray and optical spectroscopy of $\pi$ Aqr}

\Received{}
\Accepted{}
\Published{}
\title{X-ray and optical spectroscopic study of a $\gamma$ Cassiopeiae analog source $\pi$ Aquarii}
\author{
Masahiro~\textsc{Tsujimoto},\altaffilmark{1}
Takayuki~\textsc{Hayashi},\altaffilmark{2}
Kumiko~\textsc{Morihana},\altaffilmark{3}
Yuki~\textsc{Moritani}\altaffilmark{3}
}
\altaffiltext{1}{Japan Aerospace Exploration Agency, Institute of Space and Astronautical Science, Chuo-ku, Sagamihara, Kanagawa 252-5210, Japan}
\email{tsujimot@astro.isas.jaxa.jp}
\altaffiltext{2}{NASA's Goddard Space Flight Center, Greenbelt, MD 20771, USA}
\altaffiltext{3}{Subaru Telescope, National Astronomical Observatory of Japan, Hilo, HI 96720, USA}
\maketitle

\begin{abstract}
 $\gamma$ Cas analog sources are a subset of Be stars that emit intense and hard X-ray
 emission. Two competing ideas for their X-ray production mechanism are (a) the magnetic
 activities of the Be star and its disk and (b) the accretion from the Be star to an
 unidentified compact object. Among such sources, $\pi$ Aqr plays a pivotal role as it
 is one of the only two spectroscopic binaries observed for many orbital cycles and one
 of the three sources with X-ray brightness sufficient for detailed X-ray
 spectroscopy. \citet{Bjorkman2002} estimated the secondary mass $>$2.0~$M_{\odot}$ with
 optical spectroscopy, which would argue against the compact object being a white dwarf
 (WD). However, their dynamical mass solution is inconsistent with an evolutionary
 solution and their radial velocity measurement is inconsistent with later work by
 \citet{Naze2019}. We revisit this issue by adding a new data set with the NuSTAR X-ray
 observatory and the HIDES \'{e}chelle spectrograph. We found that the radial velocity
 amplitude is consistent with \citet{Naze2019}, which is only a half of that claimed by
 \citet{Bjorkman2002}. Fixing the radial velocity amplitude of the primary, the
 secondary mass is estimated as $<$1.4~$M_{\odot}$ over an assumed range of the primary
 mass and the inclination angle.  We further constrained the inclination angle and the
 secondary mass independently by fitting the X-ray spectra with a non-magnetic or
 magnetic accreting WD model under the assumption that the secondary is indeed a WD. The
 two results match well. We thus argue that the possibility of the secondary being a WD
 should not be excluded for $\pi$ Aqr.
\end{abstract}

%\linenumbers

%%%%%%%%%%%%%%%%%%%%%%%%%%%%%%%%%%%%%%%%%%%%%%%%%%%%%%%%%%%%
\section{Introduction}\label{s1}
%%%%%%%%%%%%%%%%%%%%%%%%%%%%%%%%%%%%%%%%%%%%%%%%%%%%%%%%%%%%
$\gamma$ Cas analog sources are a subset of Be stars that emit intense and hard X-ray
emission distinctive from other Be-type stars \citep{Berghofer1997a}. The X-ray
luminosity is 10$^{32}$--10$^{33}$~erg~s$^{-1}$ and the emission extends beyond
10~keV. The cause for such X-ray emission remains a puzzle for more than four
decades. Two competing ideas for producing the X-ray emitting plasma are (a) the
magnetic activities of the Be star and its disk \citep{robinson00,Smith2004} and (b) the
accretion from the Be star to an unidentified compact object. The compact object is
either a white dwarf (WD; \cite{haberl95,kubo98,hamaguchi16}) or a neutron star (NS;
\cite{white82,Postnov2017}). More details can be found in \citet{smith16} for the review
and in a conference site in 2018\footnote{Materials available at
\url{https://gammacas-enigma.sciencesconf.org/}.} for more recent developments and
related topics.

Since the X-ray detection from the prototypical $\gamma$ Cas \citep{mason76}, the number
of $\gamma$ Cas analog sources has increased to a few dozens thanks mostly to the recent
X-ray surveys using XMM-Newton observatory \citep{motch10,Naze2020b}. The increase will
be accelerated with eROSITA \citep{Merloni2012a} in the near future.  $\pi$ Aqr is one
of the recent addition to the group \citep{naze2017}. This source may play a pivotal role
in solving the long-standing puzzle of the X-ray production mechanism of $\gamma$ Cas
analog sources for its unique features. First, it is one of a few, along with $\gamma$
Cas and HD\,110432 (BZ Cru), that are bright enough in X-ray flux to allow detailed
X-ray spectroscopy with modern instruments. Second, it is one of the two, along with
$\gamma$ Cas, that are known to be a binary with optical spectroscopy covering many
orbital cycles. Note that \citet{naze2022} revealed the spectroscopic binary nature of
six more $\gamma$ Cas analogs. Third, the source was observed both in the normal B and
Be phases. It is common that Be stars make transitions between the two phases due to the
formation and the destruction of the circumstellar disks and shells, but $\pi$ Aqr is
the only one that exhibited this transition among $\gamma$ Cas analog sources to date.

\citet{Bjorkman2002} discovered the spectroscopic binary nature of this source. They
monitored the H$\alpha$ profile between 1996 and 2001 when $\pi$ Aqr was in the normal B
star phase. Both emission and absorption lines were detected, which changed their radial
velocity sinusoidally in the opposite phase. They argued that the absorption is from the
primary B star and the emission is from the unknown secondary. The claimed mass for the
primary and the secondary are, respectively, $M_1 \sin{i}=12.4~M_{\odot}$ and $M_2
\sin{i}=2.0~M_{\odot}$, where $i$ is the orbital inclination angle. This apparently
excludes the possibility that the secondary is a WD beyond the Chandrasekhar mass limit
of 1.38~$M_{\odot}$. However, the following studies \citep{Zharikov2013,Naze2019} pointed
out some inconsistencies in this orbital solution.

The purpose of this paper is to revisit the mass estimate of the secondary with new sets
of data both in the X-ray and optical spectroscopy. For $\gamma$ Cas and HD\,110432, it
was demonstrated that the secondary mass can be constrained, independently from optical
measurements, by fitting the broad-band (0.5--50~keV) X-ray spectra with a physically
motivated model if the WD scenario is adopted as a working hypothesis
\citep{Tsujimoto2018}. The two independent mass estimates by the optical and X-ray
spectroscopy were consistent with each other for $\gamma$ Cas, which is a spectroscopic
binary. We apply this approach to $\pi$ Aqr.

A comprehensive study of $\pi$ Aqr was made by \citet{Naze2019} with a year-long
monitoring in the 2018 season using both the Neil Gehrels Swift Observatory in the
X-rays below 10~keV and the \'{e}chelle spectrograph mounted at the 1.2~m TIGRE
telescope in the optical. We supplement their work in the 2019 season by adding the
data with the NuSTAR observatory in the X-rays above 10~keV and the \'{e}chelle
spectrograph mounted at the 1.88~m telescope in the Okayama Astrophysical Observatory
(OAO).

\medskip

The outline of this paper is as follows. In \S~\ref{s2}, we present the NuSTAR X-ray
(\S~\ref{s2-1}) and the OAO optical (\S~\ref{s2-3}) data sets. To complement the soft
X-ray band spectrum, we also use the archived XMM-Newton data (\S~\ref{s2-2}; the same
data presented in \cite{naze2017}). In \S~\ref{s3}, we present the analysis of the X-ray
(\S~\ref{s3-1}) and optical (\S~\ref{s3-2}) spectra separately. In \S~\ref{s4}, we
revisit the secondary mass estimate based on the literature and examine consistency with
our result. We will argue for a lower mass estimate for the secondary than was
originally proposed.

\medskip

We adopt the following parameters for $\pi$ Aqr. The distance is 286~pc based on a Gaia
parallax measurement \citep{Collaboration2016a,Brown2018}. The binary orbit is 84.1~day
with the orbital phase $\phi_{\mathrm{orb}}=0$ at the heliocentric Julian date of
2450275.5. The ephemeris was derived spectroscopically in 1996--2000 \citep{Bjorkman2002}
and was confirmed stable in 2018 \citep{Naze2019}. The effective temperature of the
primary is $T_{\mathrm{eff}}=24 \pm 1$~kK and the visual extinction is $A_{V}=0.15$~mag
based on the SED during a normal B star phase \citep{Bjorkman2002}.

%%%%%%%%%%%%%%%%%%%%%%%%%%%%%%%%%%%%%%%%%%%%%%%%%%%%%%%%%%%%
\section{Observations and Data Reduction}\label{s2}
%%%%%%%%%%%%%%%%%%%%%%%%%%%%%%%%%%%%%%%%%%%%%%%%%%%%%%%%%%%%
\subsection{NuSTAR}\label{s2-1}
We observed $\pi$ Aqr with the Nuclear Spectroscopic Telescope Array (NuSTAR;
\cite{harrison2013}) on 2019 November 4--5 for a 54~ks exposure with an observation
identification number 30501009002. NuSTAR is a space-born X-ray telescope with a
wide-band coverage in the 3--79~keV band. Above $\sim$10~keV, in particular, the
telescope has better sensitivity than previous telescopes by two orders of magnitude
thanks to the focusing mirrors employed in this energy band for the first time.

Two focal plane module (FPM) instruments are placed at the focus of two telescopes,
which are called FPMA and FPMB. Each module carries an array of CdZnTe pixel detectors
surrounded by the CsI anti-coincidence detectors. The array covers a 12\farcm0 field
with a telescope half-power diameter of 1\farcm0. The energy resolution is modest of
400-900~eV in the 10--70~keV range. These features make this telescope uniquely suited
for our present study, in which X-ray spectral curvature above 10~keV is a key to
understanding the nature of the source.

We started from the X-ray cleaned event list produced by the standard pipeline tool
\texttt{nupipeline} version 0.4.6. No other X-ray source was found in the image. We
extracted source events from a circle of a 3\arcmin\ radius and background events from
an annulus of 4\arcmin--6\arcmin\ concentric to the source. The resultant source and
background (normalized to the extraction area) count rates were 1.1 and 0.16~s$^{-1}$
respectively when the two instruments are combined.

\begin{figure}[htbp]
 \begin{center}
  \includegraphics[width=1.0\columnwidth, bb=0 0 529 379, clip]{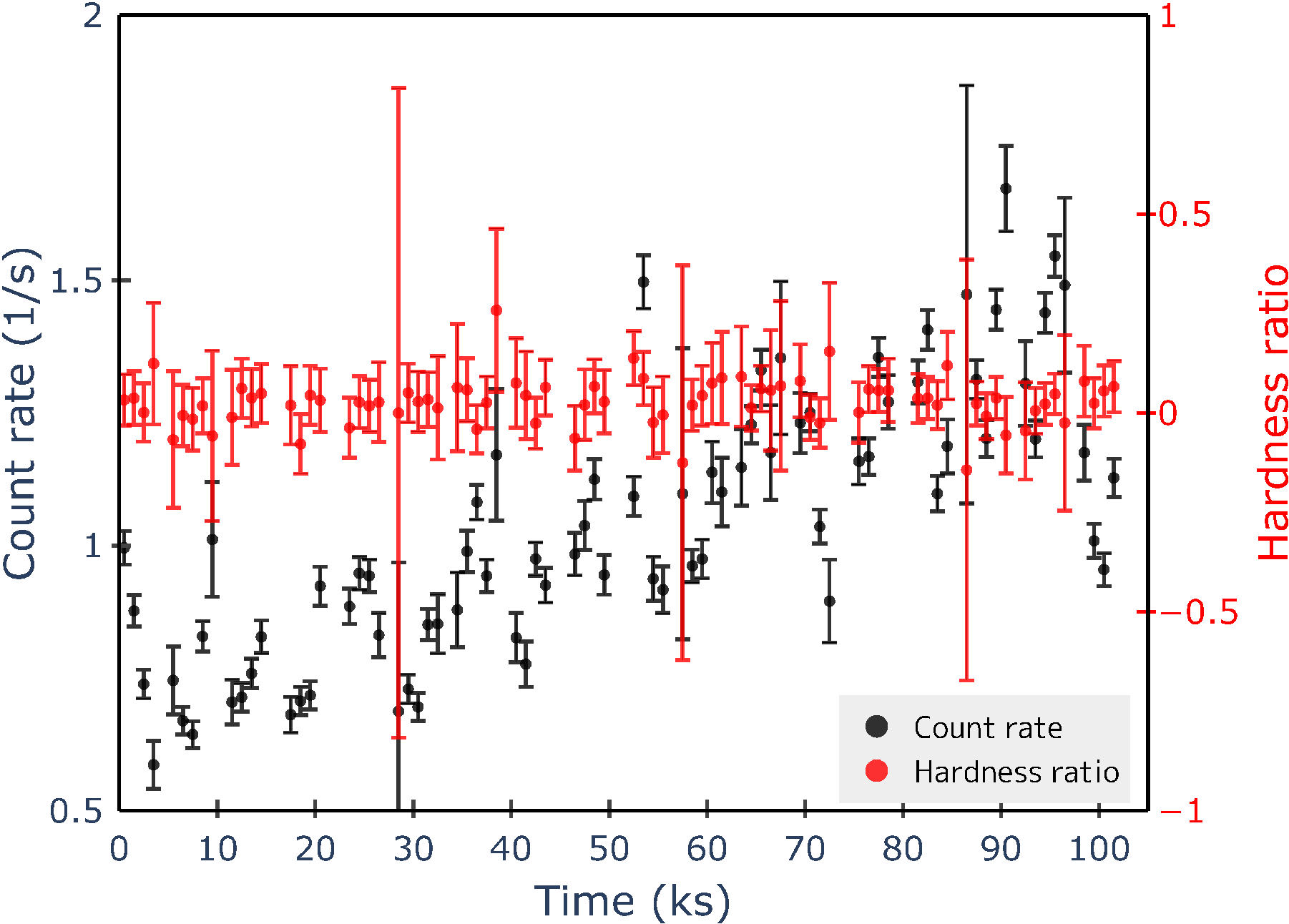}
 \end{center} 
 \caption{Count rate (1.4--76.3~keV) and hardness ratio ($(H-S)/(H+S)$, where $S$ and $H$
 are count rates of 1.4--5.4 and 5.4--76.2~keV, respectively) binned by 1~ks of the
 NuSTAR FPMA and FPMB combined.}
 \label{f02}
\end{figure}

The light curve is shown in figure~\ref{f02}. On one hand, the count rate of events in
the 1.4--76.3~keV band shows an increasing trend with some variability throughout the
observation. On the other hand, the spectral hardness defined by $(H-S)/(H+S)$ is
constant, where $H$ and $S$ are the count rates hard- and soft-ward of the median energy
of all events (5.4~keV). We thus consider that there is no change in the spectral shape
besides the flux and construct spectrum without dividing by time.

\begin{figure}[htbp]
 \begin{center}
 \includegraphics[width=1.0\columnwidth, bb=0 0 529 379, clip]{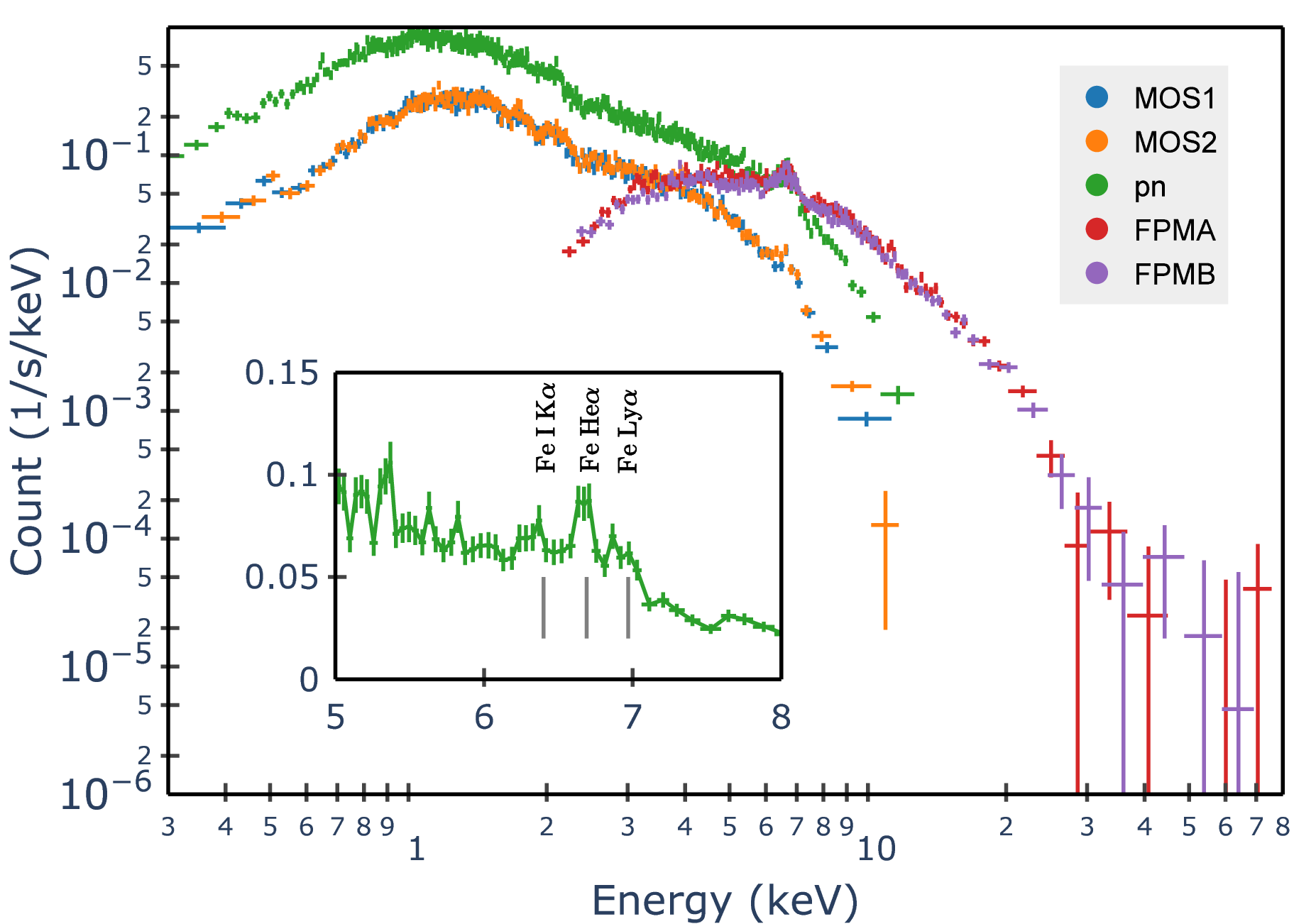}
 \end{center} 
 \caption{Spectra of EPIC MOS1, MOS2, pn and NuSTAR FPMA and FPMB. The inset gives an
 enlarged view in the Fe K band of the EPIC pn spectrum.}
 \label{f01}
\end{figure}

We made background subtracted spectra separately for FPMA and FPMB. The auxiliary
response files (ARF) of the telescope as well as the redistribution matrix functions
(RMF) of the instruments were produced by the \texttt{nuproducts} version
om.3.0. Figure~\ref{f01} shows the spectra.

\subsection{XMM-Newton}\label{s2-2}
We retrieved the archival data of $\pi$ Aqr observed with XMM-Newton. This is to
complement the X-ray data down to 0.3~keV and to have a better energy resolution than
NuSTAR at the Fe K band in the 6--7~keV to resolve the prominent three emission lines of
the Fe XXV Ly$\alpha$ and Fe XXIV He$\alpha$ lines from high-temperature plasma as well
as the Fe fluorescence line as a reprocessed product of it. The observation with an
identification number 0720390701 was made on 2013 November 16--17 for a 52~ks
exposure. The details of the observations and the results can be found in \citet{naze2017}.

We used the data taken with the European Photon Imaging Camera (EPIC) equipped with two
MOS cameras \citep{Turner2001} and one pn camera \citep{Struder2001}. After the
standard pipeline processing using the \texttt{emchain} and \texttt{epchain} tools in
the \texttt{SAS} package version 17.0, we removed events taken during high background or
those having a pixel distribution pattern of non X-ray events. We extracted source
events from a circle of a 30\arcsec\ radius and background events from an annulus of
75\arcsec--90\arcsec\ concentric to the source.

As described in \citet{naze2017}, the count rate varied by a factor of $\sim$3 while the
hardness ratio was constant, similarly to the NuSTAR data (\S~\ref{s2-1}). We thus
constructed spectra by integrating the entire observation. The ARF and RMF were generated
using the standard tools in the \texttt{SAS} package. Figure~\ref{f01} shows the
spectra.

\subsection{OAO}\label{s2-3}
We observed $\pi$ Aqr 13 times from 2019-09-06 to 2019-12-23 using the HIgh Dispersion
Echelle Spectrograph (HIDES) placed at the coud\'{e} focus of the 188~cm telescope in
the Okayama Astronomical Observatory (OAO). The OAO is no longer supported by
the National Astronomical Observatory of Japan (NAOJ) as a common-use facility open to
the community, but it continues operation under the contract among NAOJ, Tokyo Institute
of Technology, and Asakuchi City.
HIDES is an \'{e}chelle spectrograph equipped with a mosaic of three 2k$\times$4k CCD
cameras covering a wide range of 3800--7500~\AA. A fiber link is used between the
telescope and HIDES to improve optical throughput while keeping the radial velocity
precision to a few km~s$^{-1}$ at a resolution of $>$50,000 \citep{Kambe2013}.

\begin{figure}[htbp]
 \begin{center}
 \includegraphics[width=1.0\columnwidth, bb=0 0 529 379, clip]{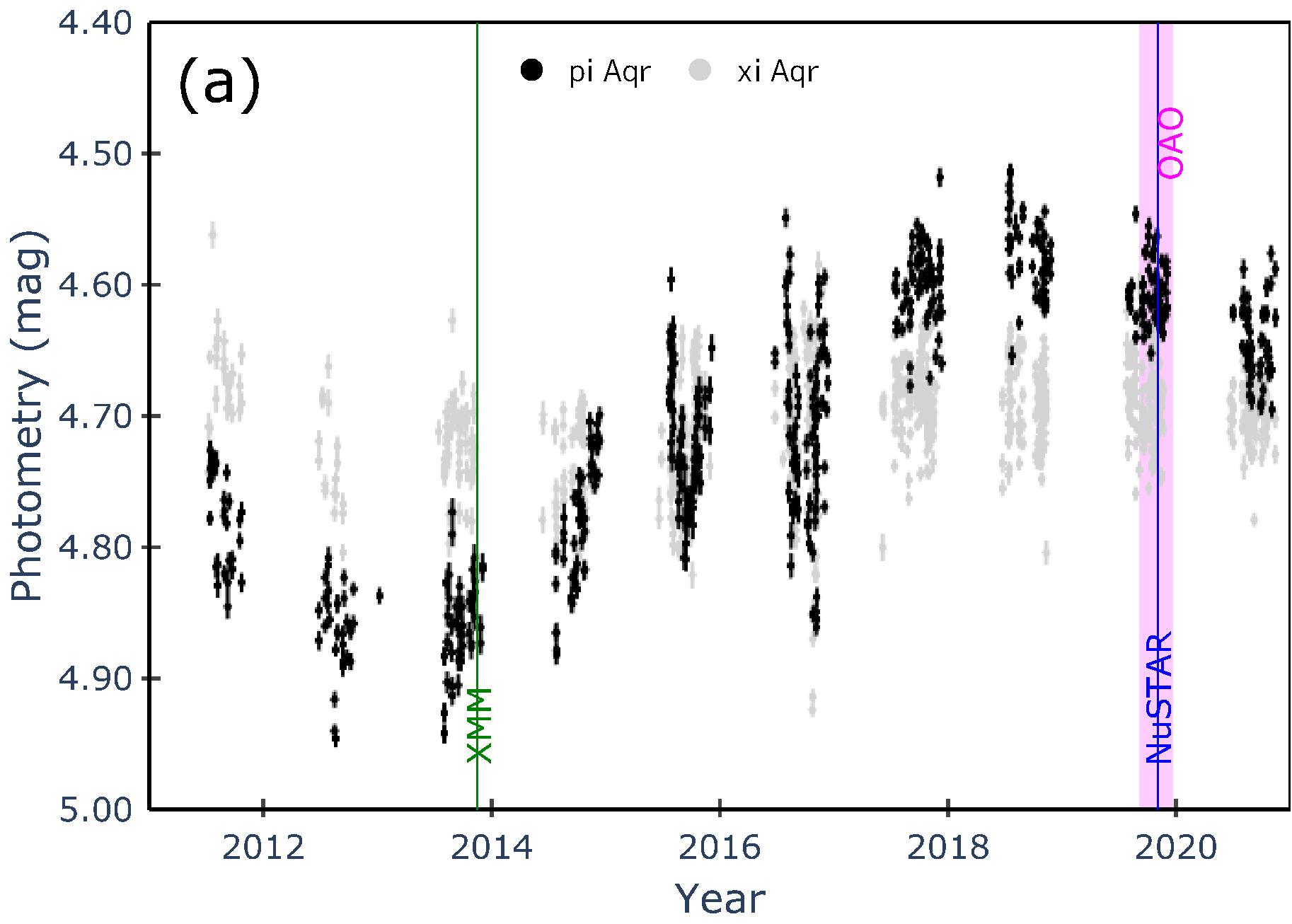}
 \includegraphics[width=1.0\columnwidth, bb=0 0 529 379, clip]{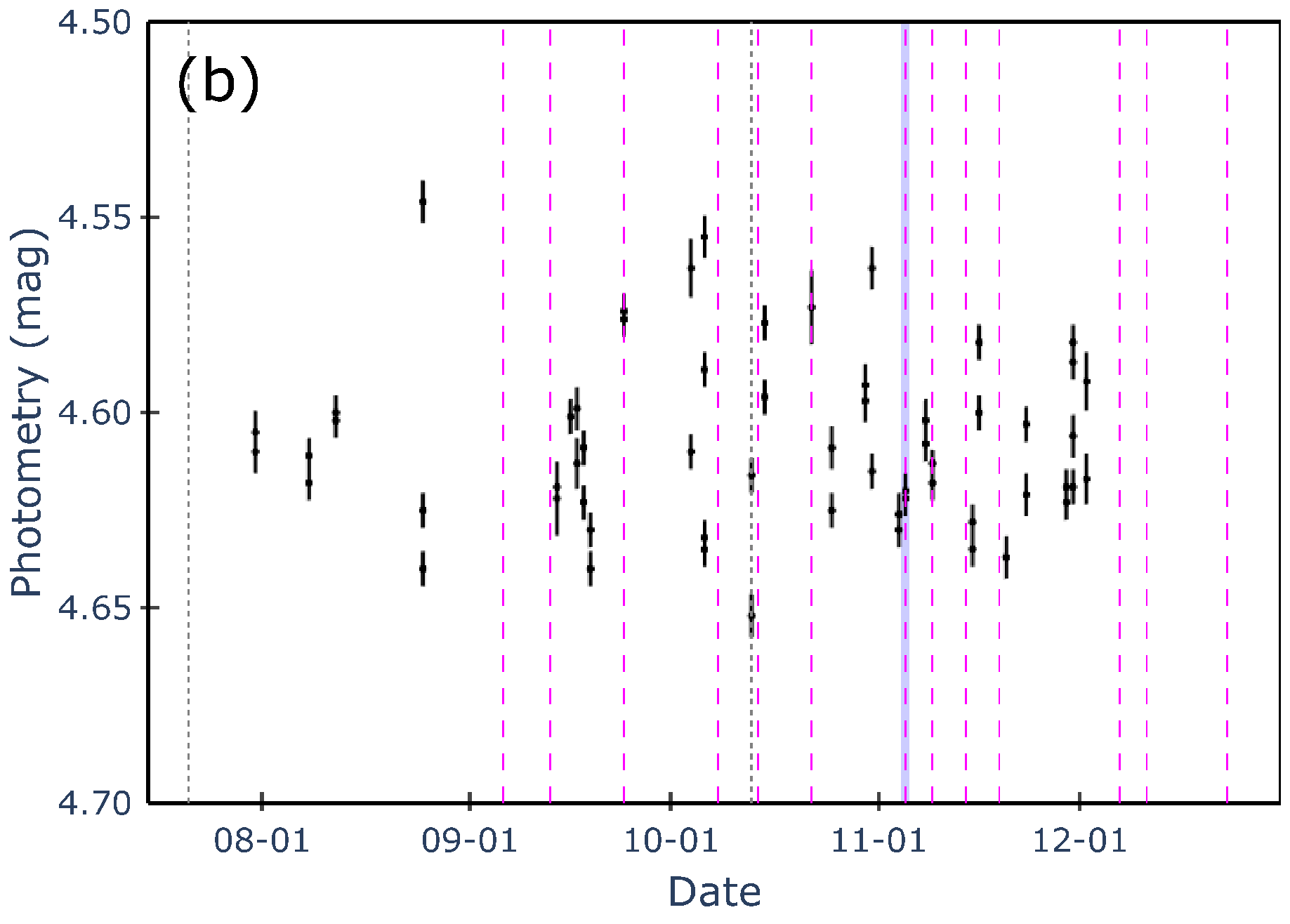}
 \end{center} 
 \caption{(a) Decadal $V$-band light curve of $\pi$ Aqr and its near-by A0V star $\xi$
 Aqr for reference taken with the Kamogata/Kiso/Kyoto Wide-field Survey
 \citep{Maehara2013a}. The epochs of the data presented in this paper is given in
 colors. (b) Close-up view of the 2019 season. The epochs of each OAO observation and
 the duration of the NuSTAR observation are shown with colors. Two dashed lines indicate
 the epoch of $\phi_{\mathrm{orb}}=0$ based on the ephemeris by \citet{Bjorkman2002}.}
 \label{f04}
\end{figure}

Table~\ref{t01} shows the observation log. The planned cadence was once per week, which
was randomized to some extent due to weather and facility issues. Figure~\ref{f04} (a)
shows the light curve for the last ten years taken with an automated sky monitor
\citep{Maehara2013a}. $\pi$ Aqr exhibits a long-term variability in this time scale in
comparison to a nearby field star $\xi$ Aqr. During our OAO observation campaign, the
star was relatively stable in the $V$-band magnitude. Figure~\ref{f04} (b) shows a
close-up view of the 2019 season. Variability within the season is presumably due to
systematics as similar variation is also seen in the reference star. A total of 13 OAO
observations cover 1.28 orbital cycles of $\pi$ Aqr. The seventh observation was
executed during the NuSTAR observation unintentionally due to unrelated schedule changes
in both observatories.

Each observation has a mean exposure time and air mass respectively of 18.5~min and
1.4. We took bias and flat data before and after observations every night, and
comparison data with a ThAr lamp.
We used pipeline products, in which the bias was subtracted, the \'{e}chelle orders
were traced, scattering and cosmic-rays were removed, the image flatness was corrected,
one-dimensional spectra were extracted, wavelengths were registered, and continuum was
corrected for flatness. Between the 10'th and 11'th observations, a small earthquake of
a magnitude of 3 hit the observatory on 2019 November 26, which made some shifts in the
optical alignment of the instrument. The \'{e}chelle orders changed for some features as
a result, but no systematic difference was found in the registered wavelengths. The
typical uncertainty of the wavelength is estimated as $\sim$0.03~\AA\ from telluric
features.

\begin{table}[hbtp]
 \caption{Observation log}\label{t01}
 \begin{tabular}{cccccc}
  \hline
  \hline
  Num & Date & Time\footnotemark[$*$] & $t_{\mathrm{exp}}$\footnotemark[$\dagger$] & Air
		  mass\footnotemark[$*$] & $\phi_{\mathrm{orb}}$\footnotemark[$\ddagger$] \\
      &      & (UT) & (s)                                        & & \\ 
  \hline
  1 & 2019-09-06 & 11:25:08 & 1115 & 1.72 & 0.56 \\
  2 & 2019-09-13 & 10:52:54 & 1128 & 1.76 & 0.65 \\
  3 & 2019-09-24 & 16:45:51 & 1200 & 1.91 & 0.78 \\
  4 & 2019-10-08 & 10:06:07 & 1166 & 1.45 & 0.95 \\
  5 & 2019-10-14 & 10:32:55 & 1098 & 1.29 & 0.02 \\
  6 & 2019-10-22 & 12:27:00 & 1065 & 1.23 & 0.11 \\
  7\footnotemark[$\S$] & 2019-11-05 & 09:11:50 & 1095 & 1.28 & 0.28 \\
  8 & 2019-11-09 & 09:24:16 & 1076 & 1.23 & 0.33 \\
  9 & 2019-11-14 & 08:51:32 & 1083 & 1.25 & 0.39 \\
  10& 2019-11-19 & 09:11:23 & 1064 & 1.20 & 0.44 \\
  11& 2019-12-07 & 08:44:33 & 1058 & 1.20 & 0.66 \\
  12& 2019-12-11 & 08:49:17 & 1060 & 1.21 & 0.71 \\
  13& 2019-12-23 & 10:43:29 & 1203 & 1.82 & 0.85 \\
  \hline
  \multicolumn{5}{@{}l@{}}{\hbox to 0pt{\parbox{80mm}{
  \footnotesize
  \par\noindent
  \footnotemark[$*$] Mean value in the exposure.
  \par\noindent
  \footnotemark[$\dagger$] Exposure time.
  \par\noindent
  \footnotemark[$\ddagger$] Orbital phase based on the ephemeris by \citet{Bjorkman2002}.
  \par\noindent
  \footnotemark[$\S$] Simultaneous with the NuSTAR observation.
 \par\noindent
 }\hss}
 }
\end{tabular}
\end{table}

%%%%%%%%%%%%%%%%%%%%%%%%%%%%%%%%%%%%%%%%%%%%%%%%%%%%%%%%%%%%
\section{Analysis}\label{s3}
%%%%%%%%%%%%%%%%%%%%%%%%%%%%%%%%%%%%%%%%%%%%%%%%%%%%%%%%%%%%
\subsection{X-ray spectra}\label{s3-1}
Figure~\ref{f01} shows the spectra of XMM-Newton EPIC and NuSTAR FPM. As commonly seen
in all $\gamma$ Cas analog sources, the X-ray spectrum of $\pi$ Aqr is also
characterized by a hard continuum extending up to 50~keV, and the Fe K complex with the
quasi-neutral Fe\emissiontype{I} K$\alpha$ line at 6.4~keV as well as the highly-ionized
Fe\emissiontype{XXV} He$\alpha$ line at 6.7~keV and Fe\emissiontype{XXVI} Ly$\alpha$
line at 7.0~keV as shown in the inset.

We apply two accreting WD models ---non-magnetic (\S~\ref{s3-1-1}) and magnetic
(\S~\ref{s3-1-2})--- to $\pi$ Aqr as was done for $\gamma$ Cas and HD\,110432
\citep{Tsujimoto2018}. We have no clue if $\pi$ Aqr, or any $\gamma$ Cas analog sources,
host a non-magnetic or magnetic WD, thus we need to test both possibilities, which
require different models.

All the XMM-Newton EPIC and the NuSTAR FPM spectra
(figure~\ref{f02}) were fitted simultaneously. The source is known to be variable in the
X-rays through a year-long monitor using the Neil Gehrels Swift Observatory
\citep{Naze2019}, hence we allowed the normalization of the NuSTAR spectra to be
different from that of the XMM-Newton spectra. The hydrogen-equivalent column density
($N_{\mathrm{H}}$) also varies \citep{Naze2019}, but because the extinction is too low
to affect the hard band X-ray spectra of the NuSTAR data, we kept it to be the same
between the XMM-Newton and NuSTAR models. The value of $N_{\mathrm{H}}$ due to the
interstellar matter is fixed to 3.16$\times$10$^{21}$~cm$^{-2}$ derived from two
consistent measurements of $A_{V}=0.15$~mag \citep{Bjorkman2002} and $E(B-V)=0.07$~mag
\citep{Jenkins2009} and an additional local extinction was derived by the fitting.

\subsubsection{Non-magnetic accreting WD model}\label{s3-1-1}
Non-magnetic accreting WDs have magnetic fields ($<$0.1~MG) that are not strong enough to
disturb the accretion from the companion star. They are known as copious X-ray sources
\citep{mukai17}. The X-ray emission comes from the boundary layer \citep{patterson85},
in which the gravitational energy released through accretion is dissipated into heat by
the friction with the slowly rotating surface of the WD.

The X-ray spectra at quiescence are well modeled by the cooling flow model
\citep{pandel05,wada17}, which is a convolution of the single temperature plasma at a
collisional ionization equilibrium and the isobaric temperature gradient. In addition,
non-magnetic accreting WDs often show partial covering absorber local to the star and
the fluorescent and scattered emission reprocessed at the WD surface and the accretion
disk.

We fitted the observed spectra first with the \texttt{mkcflow} model \citep{mushotzky88}
attenuated by the \texttt{tbabs} model \citep{wilms00} for the interstellar
absorption. The free parameters are the maximum temperature of the temperature
gradient ($T_{\mathrm{max}}$), the mass accretion rate ($\dot{M}_{X}$), and the
metallicity of the plasma ($Z$) relative to the ISM value \citep{wilms00}. We fixed the
minimum temperature of the cooling flow to the lowest available value (80~eV). We also
made the NuSTAR flux relative to XMM-Newton as a free parameter.

\begin{figure}[!hbtp]
 \begin{center}
 \includegraphics[width=1.0\columnwidth,bb=0 0 522 371,clip]{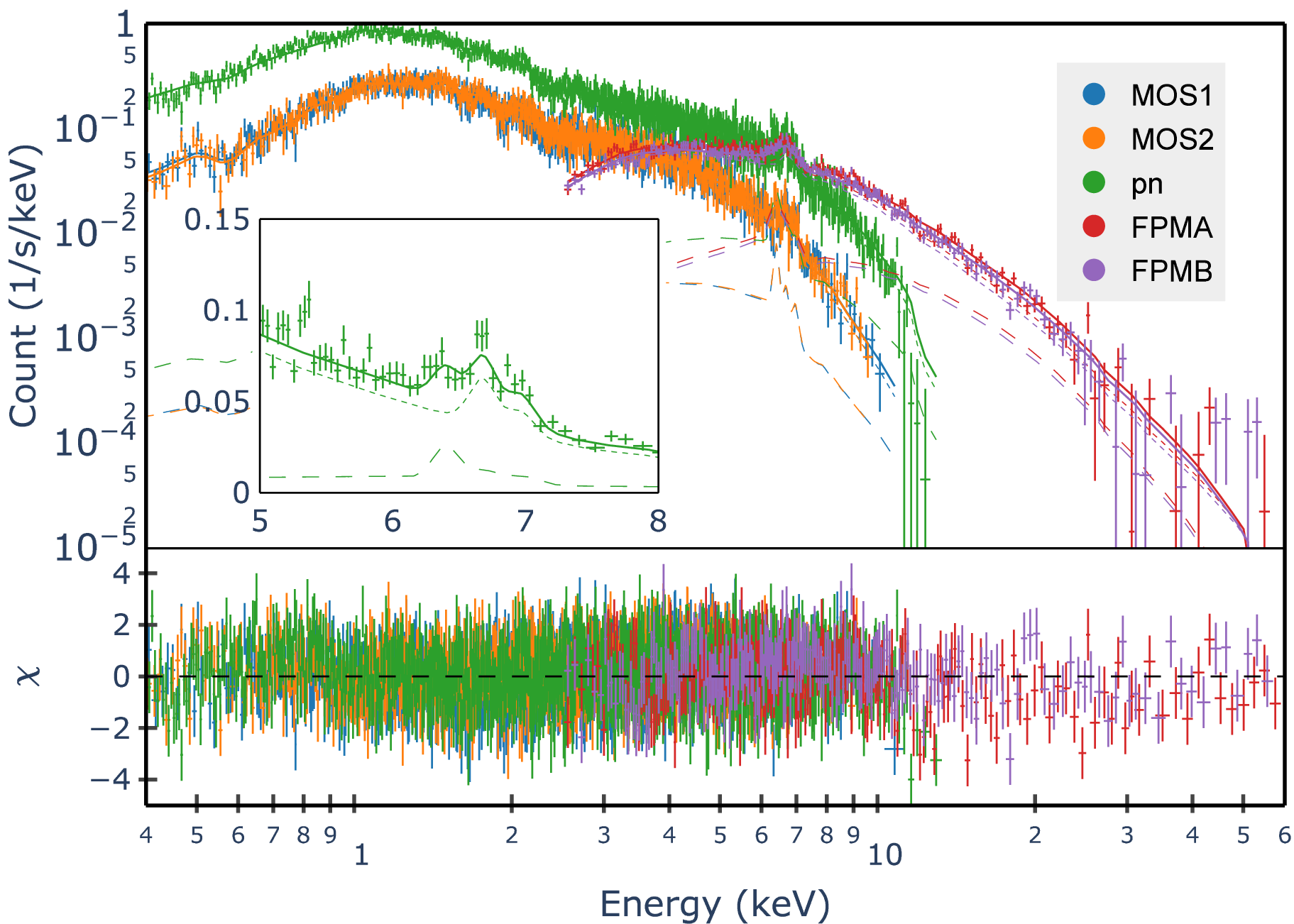}
 \end{center}
 \caption{Spectrum and the best-fit non-magnetic accreting WD model. The
 upper panels show the data (plus marks) and the model (dotted, dashed, and solid curves
 respectively for the direct, reprocessed, and total components), whereas the bottom
 panel shows the residuals to the fit. The inset gives an enlarged view in the Fe K band
 of the EPIC pn spectrum.}
 \label{f08}
\end{figure}

This yielded an unsatisfactory fit with large residuals at (a) the soft-band end, (b) the
hard-band end, and (c) 6.4~keV. To account for these deviations, we modified the model
by (a) convolving with a partial covering model and (b) the Compton reflection model
\citep{magdziarz95} and by (c) adding a Gaussian model. The additional free parameters
are (a) absorption column ($N_{\mathrm{H}}$) and the covering fraction in the partial
covering model, (b) the subtended angle for the Compton reflection ($d\Omega/2\pi$), and
(c) the intensity of the 6.4~keV line represented by a Gaussian distribution with the
center and the intrinsic width fixed at 6.40~keV and 0~eV. The fluorescence and the
Compton scattering are two competing processes in the reprocessed emission under the
same geometry. No model has been developed to explain both, thus we took the following
approach. Based on \citet{george91}, we derived $d\Omega/2\pi$ from the equivalent width
(EW$_{\mathrm{Fe}}$) of the 6.4~keV line and compared it to the one derived for the
Compton scattering. We iterated the fitting until the two converged for an assumed
reflection angle $i_{r}$. We obtained a reasonable fitting result for several selected
values in the range $i_{r}=$50--60 degrees that cover the conceivable range (see
\S~\ref{s4-2}) with little difference in the fitting parameters other than
$d\Omega/2\pi$. Table~\ref{t02} and figure~\ref{f08} show the result for
$i_{r}=60$~degree.

\begin{table}[htbp]
 \caption{Best-fit parameters with the non-magnetic accreting WD model$^{*}$.}
 \label{t02}
 \begin{center}
  \begin{tabular}{llc}
   \hline
   \hline
   Model            & Parameter & $\pi$\,Aqr \\
   \hline
   \multicolumn{3}{l}{(Fixed values)}\\
   Distance         & $D$ (pc)& 286 \\
   Reflection angle & $i_{\mathrm{r}}$ (degree) & 60 \\
   \texttt{tbabs}   & $N_{\rm {H}}$ ($10^{21}$ cm$^{-2}$) & 3.16 \\
   \hline
   \multicolumn{3}{l}{(Fitted values$^{*}$)}\\
   \texttt{tbpcf}   & $N_{\rm {H}}$ ($10^{22}$ cm$^{-2}$) & 0.53$\pm$0.02 \\
                    & Covering fraction & 0.79$\pm$0.01 \\
   \texttt{mkcflow} & $T_{\mathrm{max}}$ (keV) & 19.1$_{-0.37}^{+0.29}$\\
                    & $Z$ (solar) & 0.51$\pm$0.03\\
                    & $\dot{M}_{\mathrm{X}}$ (10$^{-11} M_{\odot}$~yr$^{-1}$) & 3.15$_{-0.03}^{+0.04}$\\
   \texttt{reflect} & $d\Omega$/2$\pi$ & 0.79 \\
   \texttt{gauss}   & EW$_{\mathrm{Fe}}$ (eV) & 68 $\pm$8\\
   \texttt{const}   & NuSTAR/XMM flux & 2.26 $\pm$ 0.02\\
   \hline
   \multicolumn{2}{l}{$\chi^{2}_{\mathrm{red}}$ (d.o.f.)} & 1.04 (3350) \\
   \hline
   \multicolumn{3}{l}{\parbox{80mm}{
   \footnotesize
   \par \noindent
   \footnotemark[$*$] The errors indicate a 1$\sigma$ statistical uncertainty. \\
   }}
  \end{tabular}
 \end{center}
\end{table}

\subsubsection{Magnetic accreting WD model}\label{s3-1-2}
Next, we try fitting the spectra with a magnetic accreting WD model. In magnetic WDs,
the accretion disk is truncated and the matter accretes along the magnetic field to form
an accretion column above the magnetic poles. A strong shock is formed above the WD
surface, where the infalling kinetic energy is dissipated into heat as the X-ray
emitting plasma.

\begin{figure}[!hbtp]
 \begin{center}
 \includegraphics[width=1.0\columnwidth,bb=0 0 522 371,clip]{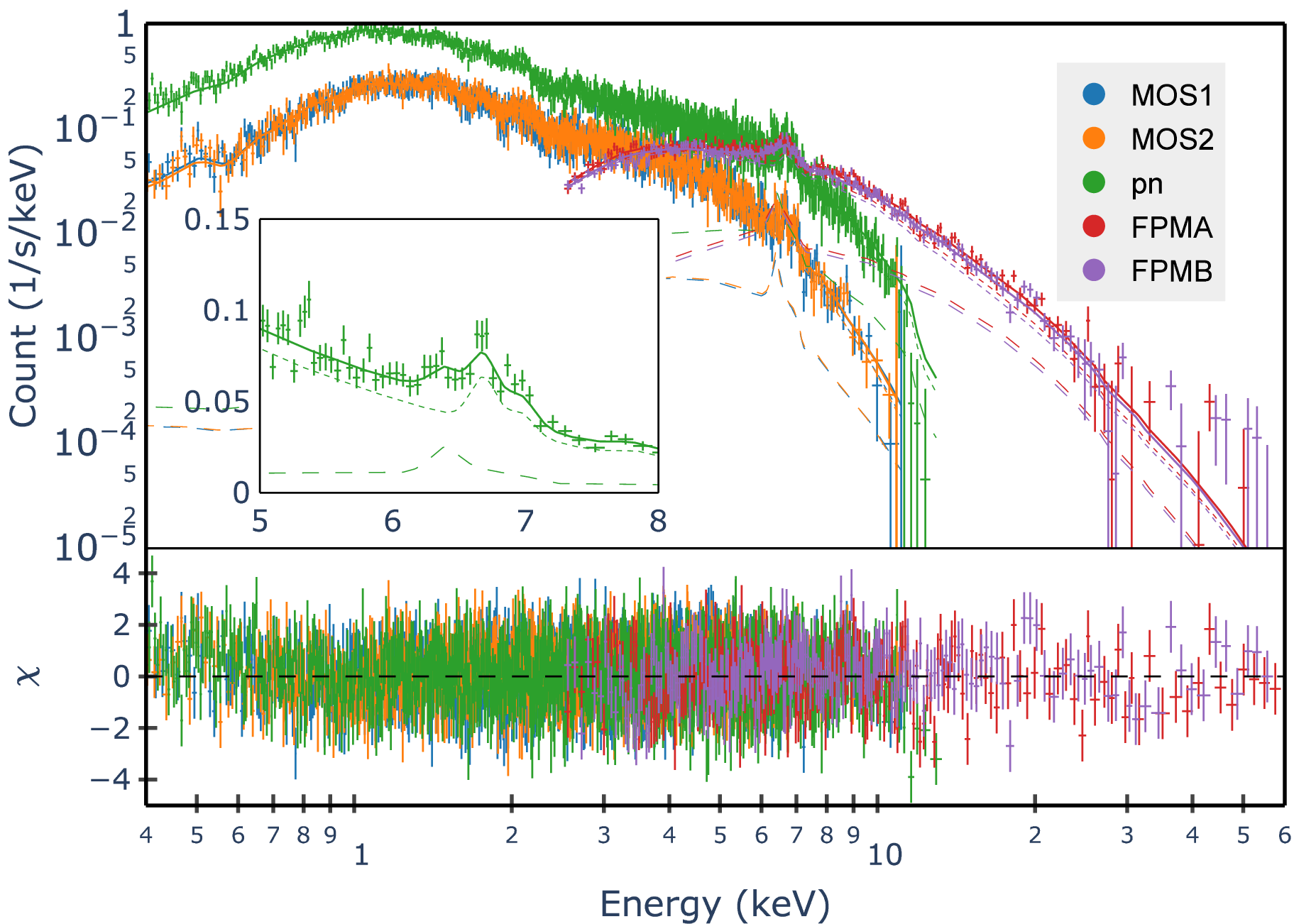}
 \end{center}
 \caption{Spectrum and the best-fit magnetic accreting WD model. The symbols follow
 figure~\ref{f08}.}
 \label{f07}
\end{figure}

X-ray spectra can be synthesized by calculating the temperature and pressure gradient
along the column and adding the emission from them. We used the \texttt{acrad} model
\citep{Hayashi2021}, in which the X-ray spectra are calculated by solving the plasma
fluid equations along the dipole magnetic field as a function of the WD mass
($M_{\mathrm{WD}}$) and the specific mass accretion rate, or the mass accretion rate per
unit area ($a$). This model further calculates the reprocessed emission. The
reprocessing takes place at the WD surface either by fluorescence or by Compton
scattering. The 6.4~keV emission line and the Compton reflection components are included
in the model as a function of the reflection angle ($i_{r}$). These parameters as well
as the metal abundance ($Z$) are constrained by the spectral fitting. Similarly to the
non-magnetic WD case (\S~\ref{s3-1-1}), we attenuated the \texttt{acrad} model with the
\texttt{tbabs} \citep{wilms00} for the interstellar absorption and \texttt{tbpcf} for
the local partial absorption. The fitting was successful with reasonable values of the
best-fit parameters (table~\ref{t03} and figure~\ref{f07}).

\begin{table}[htbp]
 \caption{Best-fit parameters with the magnetic accreting WD model$^{*}$.}
 \label{t03}
 \begin{center}
  \begin{tabular}{llc}
   \hline
   \hline
   Model            & Parameter & $\pi$ Aqr \\
   \hline
   \multicolumn{3}{l}{(Fixed values)}\\
   Distance & $D$ (pc)& 286 \\
   \texttt{tbabs} & $N_{\rm {H}}$ ($10^{21}$ cm$^{-2}$) & 3.16 \\
   \hline
   \multicolumn{3}{l}{(Fitted values$^{*}$)}\\
   \texttt{tbpcf}   & $N_{\rm {H}}$ ($10^{22}$ cm$^{-2}$) & $2.1\pm0.3$\\
                    & Covering fraction & $0.25\pm0.02$ \\
   \texttt{acrad}   & $M_{\mathrm{WD}}$ ($M_{\odot}$) & $0.51\pm0.01$  \\
                    & $Z$ (solar$^{\dagger}$) & $0.26\pm0.02$ \\
                    & $\log{a}$ (g~cm$^{-2}$~s$^{-1}$) & $3.9_{-3.8}^{+\infty}$ \\
                    & $i_{\mathrm{r}}$ (degree) & $67_{-7}^{+5}$\\
   \texttt{const}   & NuSTAR/XMM flux & $2.18\pm0.03$ \\
%   \multicolumn{3}{l}{(Derived values$^{\ddagger}$)}\\
%                    & $L_{\mathrm{X}}$ (erg~s$^{-1}$) & $1.8\times10^{32}$\\
%                    & $\dot{M}_{\mathrm{X}}$ ($M_{\odot}$~yr$^{-1}$) & \\
%                    & $f$  & $f < 2\times10^{-4}$\\
   \hline
   \multicolumn{2}{l}{$\chi^{2}_{\mathrm{red}}$ (d.o.f.)} & 1.00 (3349)\\
   \hline
   \multicolumn{3}{l}{\parbox{80mm}{
   \footnotesize
   \par \noindent
   \footnotemark[$*$] The errors indicate a 1$\sigma$ statistical uncertainty. \\
   \footnotemark[$\dagger$] Assuming that Fe represents the metals, the difference of the
   Fe abundance between the \citet{anders89} and \citet{wilms00} is corrected to match
   with the latter.\\
%   \footnotemark[$\ddagger$] $L_{\mathrm{X}}$ is derived by integrating the best-fit
%   \texttt{acrad} model in the 0.2--100~keV band. $\dot{M}_{\mathrm{X}}$ and $f$ are
%   derived using equations~\ref{e01} and \ref{e02}.\\
   }}
  \end{tabular}
 \end{center}
\end{table}

\subsection{Optical spectra}\label{s3-2}
\begin{figure*}[!hbtp]
 \begin{center}
 \includegraphics[width=1.0\textwidth,bb=0 0 1350 675,clip]{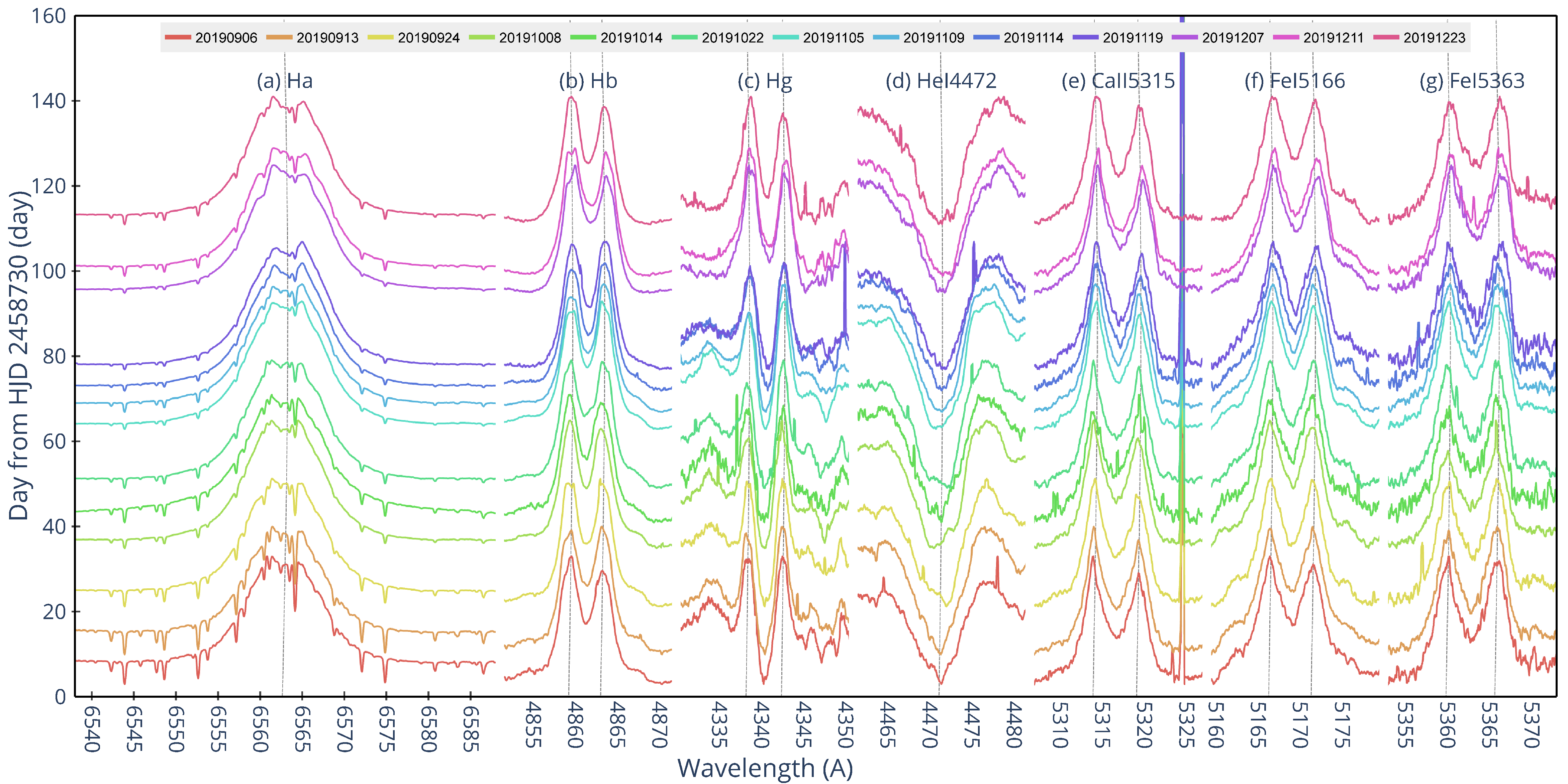}
 \end{center}
 \caption{Local spectra of some strong features shown with different colors for
 different observations. The features are detrended by subtracting the baseline
 and are normalized. The spectra are laid out along the observation dates. The width for
 H$\alpha$ is 25~\AA, while that for the others is 10~\AA. The peaks and valleys are
 traced by the dotted curve showing a fixed wavelength modulated by the Earth's motion
 around the heliocenter.}
 \label{f23}
\end{figure*}

The optical spectra have numerous features. Figure~\ref{f23} shows the local spectra of
noticeable features of H, He, and some metals laid out along the date of
observations. H$\alpha$ profile appears single-peaked emission, HeI$\lambda$4472
single-peaked absorption, and the others double-peaked emission. Fixed wavelengths
modulated by the Doppler shift of the Earth's motion around the heliocenter (--2.6 to
--28.0~km~s$^{-1}$ for the observation span) are shown with dotted curves to trace the
peaks or valleys. The peaks and valleys are fluctuating around the curve collectively
with the orbital period, suggesting that they are attributable to orbital motion.

To characterize the radial velocity changes, we used the H$\alpha$ profile, which has
the largest signal-to-noise ratio and references to compare to
\citep{Bjorkman2002,Naze2019}. We adopted the bisector method \citep{shafter1986}, which
has an advantage to avoid influences by the Be star activities \citep{Moritani2018} and
was adopted as one of the three methods in \citet{Naze2019} that yielded consistent
results. We removed the telluric absorption features and selected wavelengths of
interest in the left- and right-hand side tails of a width of 13~\AA\ to cover about
40--10\% of the peak values. Then, the left-hand side tail is reflected at a trial
center wavelength and is correlated with the right-hand side tail. The center wavelength
was derived by finding the maximum amplitude of the correlation among the trial center
wavelengths. The correlation was fitted locally with a quadrature function, yielding a
statistical error of $\lesssim$0.1~km~s$^{-1}$.

\begin{figure}[!hbtp]
 \begin{center}
 \includegraphics[width=1.0\columnwidth,bb=0 0 600 450,clip]{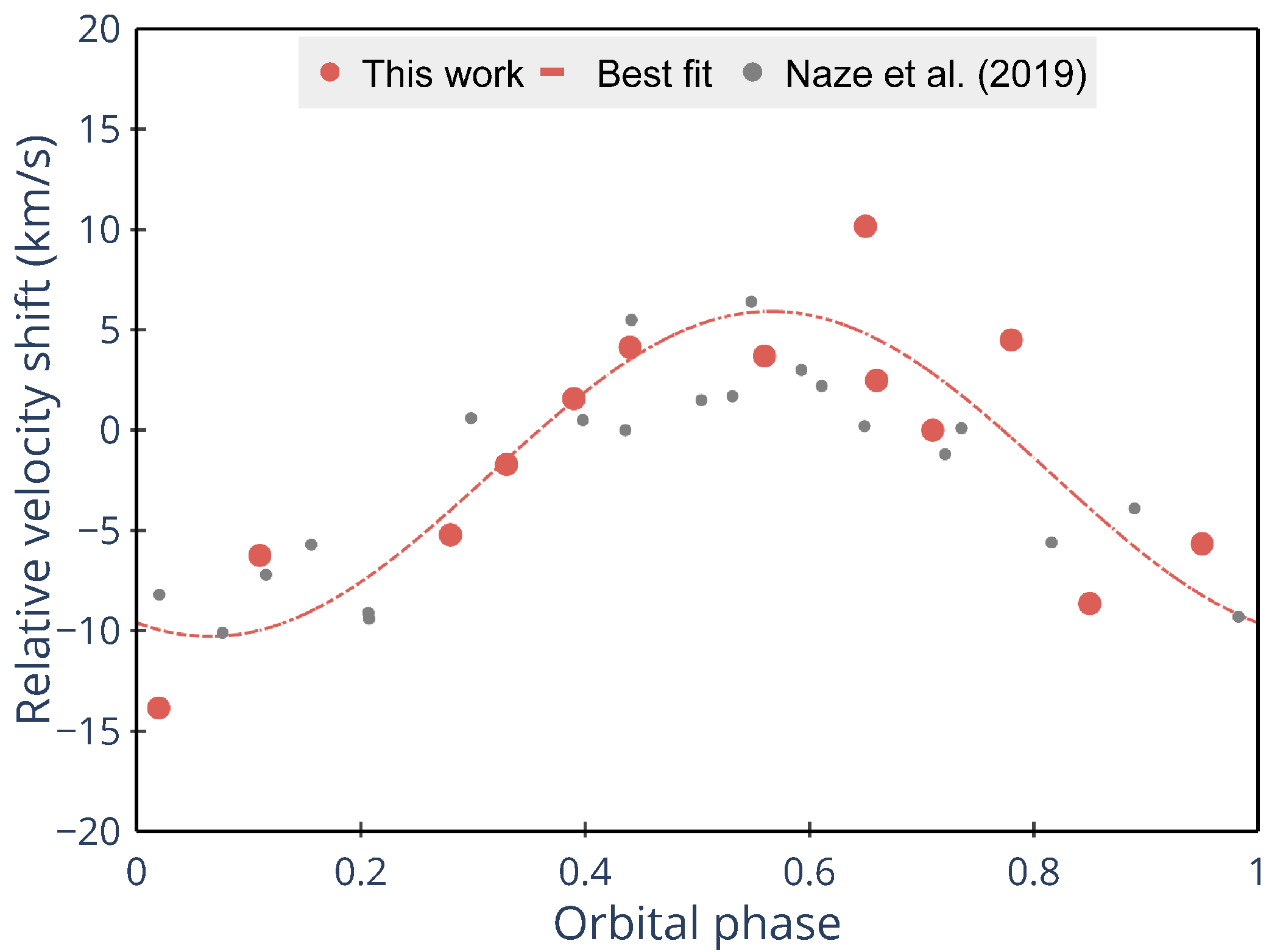}
 \end{center}
 \caption{Radial velocity relative to the mean as a function of phase. Red dots are by
 this work and red dashed curve is the best-fit sine curve. The grey dots are by
 \citet{Naze2019}.} 
 \label{f24}
\end{figure}

The result is shown in the red dots in figure~\ref{f24}, which is consistent with
\citet{Naze2019} shown in the grey dots. We fitted the red dots with a sine curve having
a fixed period of 84.1~day and constrained its amplitude of 8.1 $\pm$ 1.4~km~s$^{-1}$
and its initial phase shift of --0.4 $\pm$ 0.2 radian. As discussed in \citet{Naze2019},
the radial amplitude is less than a half of the value reported in \citet{Bjorkman2002}
as 16.7~km~s$^{-1}$.

%%%%%%%%%%%%%%%%%%%%%%%%%%%%%%%%%%%%%%%%%%%%%%%%%%%%%%%%%%%%
\section{Discussion}\label{s4}
%%%%%%%%%%%%%%%%%%%%%%%%%%%%%%%%%%%%%%%%%%%%%%%%%%%%%%%%%%%%
\subsection{Mass estimate with optical data}\label{s4-1}
We revisit the mass estimate of the binary with the latest available literature. The
mass of the binary constituents can be derived either by evolutionary or dynamical
arguments. In $\pi$ Aqr, it is known that the two mass estimates contradict each
other.

The evolutionary mass is estimated as follows. The effective temperature
($T_{\mathrm{eff}}=24\pm1$~kK) and the visual extinction ($A_{V}=$0.15~mag) are derived
from the SED fitting \citep{Bjorkman2002}. By applying the extinction and bolometric
corrections to the observed brightness and by scaling with the distance, the absolute
luminosity is obtained \citep{Zharikov2013}. We use the Gaia distance of 286~pc to
obtain $\log{L/L_{\odot}}=3.87$. Comparing $T_{\mathrm{eff}}$ and $L$ to an evolutionary
model \citep{Ekstrom2012}, the initial mass of the primary $M_1 \sim 10~M_{\odot}$ is
obtained.

The dynamical mass was estimated based on the spectra of the normal B phase
\citep{Bjorkman2002}, in which they argued that the features from both constituents were
observed; H$\alpha$ absorption attributed to the primary with $v_1\sin{i}=16.7 \pm
0.2$~km~s$^{-1}$ and the H$\alpha$ emission attributed to the secondary with
$v_2\sin{i}=101.4 \pm 2$~km~s$^{-1}$. From these, they derived $M_1\sin^{3}{i}=12.4
M_{\odot}$ and $M_2\sin^{3}{i}=2.0 M_{\odot}$ with a mass ratio of $q=M_2/M_1=0.16$.

\citet{Zharikov2013} proposed to adopt the dynamical mass over the evolutionary mass for
$M_1$; i.e., they derived $M_1=14.0 M_{\odot}$ from $M_1\sin^{3}{i}=12.4 M_{\odot}$ and
the probable inclination of $i=65-85$~degree. However, this requires $\pi$ Aqr to be
placed at a 740~pc, which is now unlikely based on the parallax distance measurement
with Gaia.

The dynamical solution is also challenged by the recent re-measurement of the radial
velocity by \citet{Naze2019}, in which they showed that $v_1\sin{i}$ is only a half of
the value by \citet{Bjorkman2002} consistently during 2014--2019. We confirmed this
independently in \S~\ref{s3-2}. \citet{Naze2019} also found no evidence of the
spectroscopic feature attributable to the secondary unlike the claim by
\citet{Bjorkman2002}. The phase of the $\pi$ Aqr was different between
\citet{Bjorkman2002} in the normal B phase and \citet{Naze2019} in the Be phase.  They
may observe different parts of the system, but the exact cause of the discrepancy is
unknown.

\begin{figure}[!hbtp]
 \begin{center}
 \includegraphics[width=1.0\columnwidth,bb=0 0 529 379,clip]{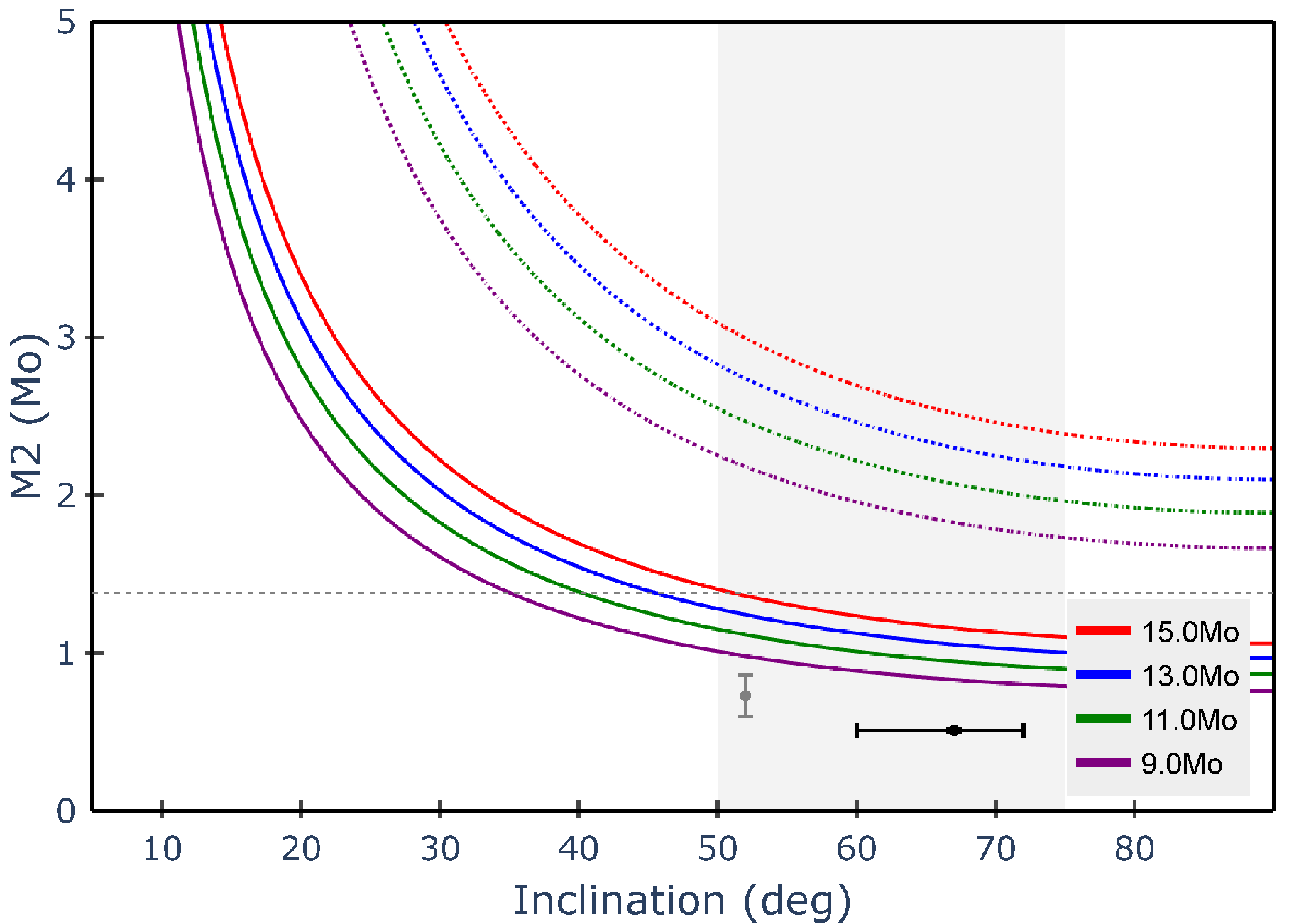}
 \end{center}
 \caption{Solution of $M_2$ as a function of an assumed inclination angle ($i$) for
 different estimates of $M_1$ (in different colors) and $v_1\sin{i}=16.7$ km~s$^{-1}$
 (dashed curves; \cite{Bjorkman2002}) and a half value of it (solid curves;
 \cite{Naze2019} and this work). The probable range of the inclination
 \citep{Bjorkman2002} is shown in the gray shaded region, while the Chandrasekhar limit for
 $M_2$ is by the dotted horizontal line. The best-fit parameters of the magnetic and non-magnetic
 accreting WD models (tables~\ref{t02} and \ref{t03}) are shown with the black and grey
 points only with statistical errors of the X-ray spectral fitting, respectively.}
 \label{f05}
\end{figure}

We thus deal with the primary mass $M_1$ as a variable in a reasonable range $9-15
M_{\odot}$ containing both the evolutionary and dynamical estimates. We deal with the
radial velocity amplitude $v_1\sin{i}$ fixed to 8.1~km~s$^{-1}$, which are confirmed by
two independent studies (\cite{Naze2019} and this work). Then, we can estimate $M_2$ for
a given inclination angle with Kepler's law (figure~\ref{f05}). \citet{Bjorkman2002}
discussed that the inclination angle should be 50--75 degree because of the large
variation of the H$\alpha$ profile, the double-peaked shape of H$\beta$ and H$\gamma$
lines (also seen in figure~\ref{f23}), and a high polarization during the brightness
maximum. The $M_2$ estimate is below the Chandrasekhar limit over an assumed range of
the primary mass and the inclination angle.

\subsection{Mass estimate with X-ray data}\label{s4-2}
We now compare the revised secondary mass estimate with the X-ray spectral fitting
result (\S~\ref{s3-1}). We should note that the application of the WD spectral models to
$\pi$Aqr is under the working hypothesis that the secondary is a WD. The assumption is
validated by the high goodness of the fitting and reasonable values of the best-fit
parameters for a WD (tables~\ref{t02} and \ref{t03}). We thus use these results.

For the magnetic WD assumption (\S~\ref{s3-1-2}), the spectral model is
physically-motivated, thus physical parameters such as $M_{\mathrm{WD}}$ are available. The model
includes the reprocessing components in itself, thus the reflection angle is also constrained as a
result of the fitting. We assume that the magnetic pole is aligned to the orbital axis
of the binary, thus the X-ray reflection angle ($i_{r}$) is the same as the orbital
inclination. The $M_{2}$ and inclination angle are thus constrained by the X-ray
spectral fitting alone, which is shown with the black point in figure~\ref{f05}.

\begin{figure}[!hbtp]
 \begin{center}
 \includegraphics[width=1.0\columnwidth,bb=0 0 595 595,clip]{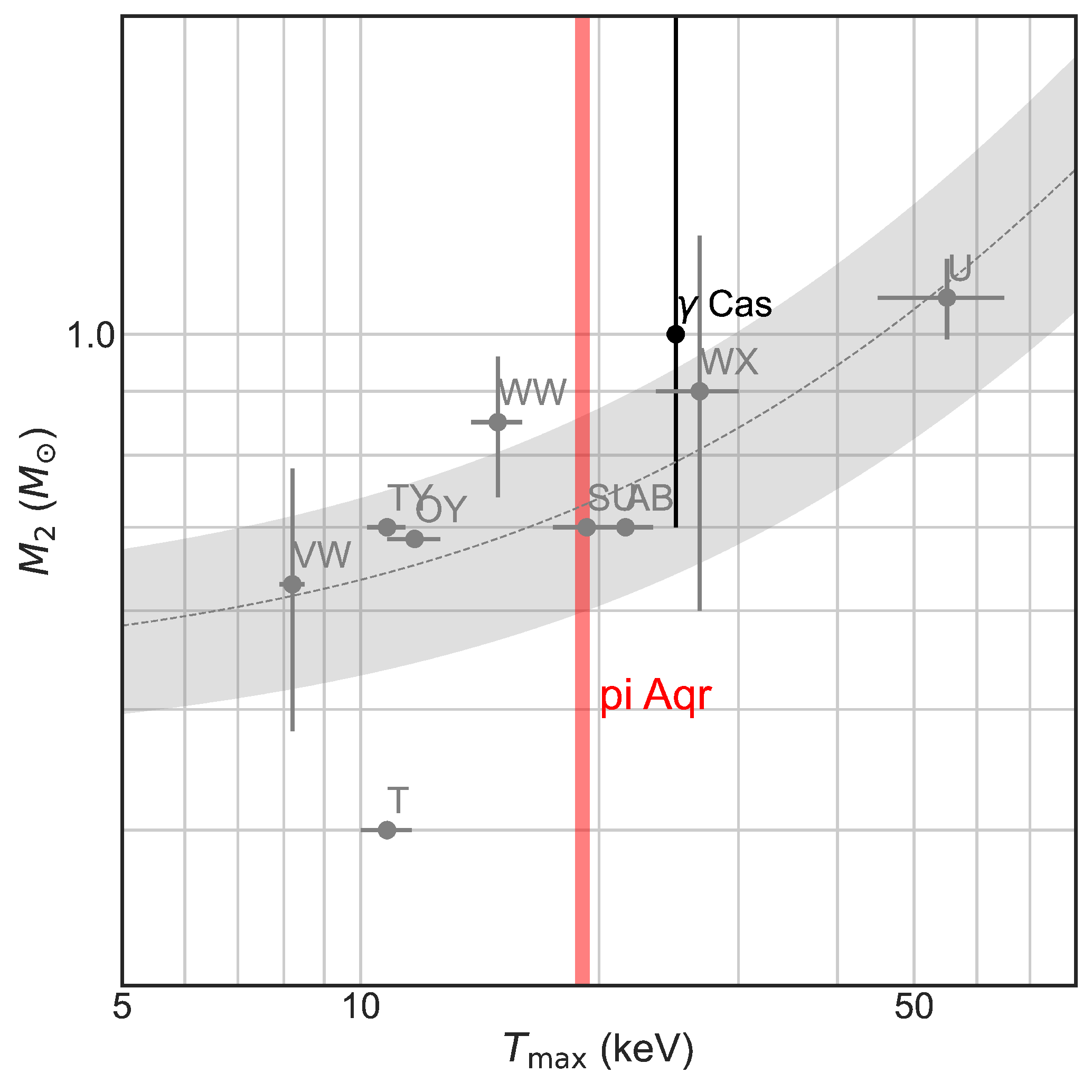}
 \end{center}
 \caption{Scatter plot of non-magnetic WD between $M_{2}$ by optical
 spectroscopy versus $T_{\mathrm{max}}$ by X-ray spectroscopy. Data are from
 \citet{pandel05} for T Leo, OY Car, VW Hyi, WX Hyi, SU UMa, TY, PsA, AB Dra, WW Cet,
 and U Gem (their abbreviations are given in the figure). For $\gamma$ Cas, the optical
 and X-ray results are respectively from \citet{harmanec00} and
 \citet{Tsujimoto2018}. For $\pi$ Aqr, the X-ray result is from table~\ref{t02} and its
 1~$\sigma$ range is shown with the red shade. The linear regression of the 
\citet{pandel05} data and its 1$\sigma$ range is shown with the dotted curve and the grey
 shade.} 
 \label{f10}
\end{figure}

For the non-magnetic WD assumption (\S~\ref{s3-1-1}), the spectral model is more
phenomenological than the one adopted for the magnetic WD assumption, thus gives less
constrained results. Still, we can compare in the $M_{2}$ versus inclination angle space
(figure~\ref{f05}). First, it is known that the maximum temperature ($T_{\mathrm{max}}$)
is correlated well with $M_{\mathrm{WD}}$ (figure~\ref{f10}). We derived the linear
regression and estimated the range for $M_{2}$ based on the $T_{\mathrm{max}}$ value
derived from the fitting to be 0.60--0.86~$M_{\odot}$. The reflection angle $i_{r}=$52
and 60 degree correspond to an inclination angle of 52 and 75~degree following a simple
geometrical argument by \citet{Tsujimoto2018}. The result for $i_{r}=$60 degree is shown
with the grey data in figure~\ref{f05}.

Whichever the assumption of magnetic or non-magnetic accreting WDs, the X-ray spectral
fitting yields a solution in the $M_{2}$ versus inclination angle space, which is closer
to the optical result adopting the smaller radial velocity amplitude of
8.1~km~s$^{-1}$. The WD mass estimate based on the X-ray spectral fitting has systematic
uncertainties, which is not shown in figure~\ref{f05}. For the magnetic model, a typical
uncertainty is $\pm$0.2~$M_{\odot}$ by comparing the spectroscopic binary mass and the
X-ray determined mass using a forerunner model (figure 8 in \cite{yuasa10}). For the
non-magnetic model, we would expect a similar systematic uncertainly from
figure~\ref{f10}. The accuracy of the X-ray determined mass in figure~\ref{f05} is
insufficient to discriminate the primary mass (9--15 $M_{\odot}$) of the revised radial
velocity of 8.1~km~s$^{-1}$, but is sufficient to exclude the secondary mass of
$>$2.0~$M_{\odot}$ previously claimed by \citet{Bjorkman2002} based on the larger radial
velocity amplitude of 16.7~km~s$^{-1}$.

This puts $\pi$ Aqr in a comparable range of $M_2$ estimate among all eight $\gamma$ Cas
analog sources with known spectroscopic binary nature. In Table 4 of \citet{naze2022},
all $\gamma$ Cas analog sources have $M_2 < 1 M_{\odot}$ except for $\pi$ Aqr having 2.4
$M_{\odot}$ based on the \citet{Bjorkman2002} solution. We showed that $M_2$ estimate
for $\pi$ Aqr should be revised to be lower. We thus conclude that the secondary of
$\pi$ Aqr being a WD is not excluded.

\subsection{Mass estimate with Gaia}\label{s4-3}
In the Gaia data release latest as of writing (DR3\footnote{See
\url{https://gea.esac.esa.int/archive/documentation/GDR3/}}), more than
8$\times$10$^{6}$ non-single stars are identified either by astrometric
\citep{halbwachs2022a} spectroscopic, or eclipsing methods.  We also retrieved the
visual binary catalog generated using the early data release of DR3
\citep{el-badry2021}, but $\pi$Aqr was not found either presumably due to the large
contrast and small separation between the binary constituents to be detected as a visual
binary.

Although $\pi$Aqr is not included in the latest Gaia's list of binary stars, the
astrometric solution is prospective. For a 10~$M_{\odot}$ and 1~$M_{\odot}$ binary with
84.1~day period at a 286~pc away, the semi-major axis of the primary is
$\sim$0.8~mas. This is within reach of the Gaia's capability \citep{holl2022a}. It is
expected that increased data volume and improved algorithms will settle the solution of
this source, along with other $\gamma$ Cas analogs, in the near future.

%%%%%%%%%%%%%%%%%%%%%%%%%%%%%%%%%%%%%%%%%%%%%%%%%%%%%%%%%%%%
\section{Summary \& Conclusion}\label{s5}
%%%%%%%%%%%%%%%%%%%%%%%%%%%%%%%%%%%%%%%%%%%%%%%%%%%%%%%%%%%%
We revisited the secondary mass estimate of $\pi$ Aqr, which is a $\gamma$ Cas analog
source. In the long-standing debate about the origin of the hard and intense X-ray
emission of the $\gamma$ Cas analog sources, $\pi$ Aqr plays an important role for being
the only two sources of known spectroscopic binary nature along with $\gamma$ Cas
covered for many orbital cycles. The previous estimate of $M_{2}$ by
\citet{Bjorkman2002} was $>2.0~M_{\odot}$, which would argue against the idea that the
X-ray is produced by accretion onto a WD. However, the mass estimate based on the
dynamical solution \citep{Bjorkman2002} is known to be inconsistent with the
evolutionary solution \citep{Zharikov2013}. We obtained new X-ray data using NuSTAR and
optical data using OAO HIDES to further investigate this discrepancy.

Our optical spectroscopy shows that the radial velocity amplitude is inconsistent with
\citet{Bjorkman2002} and is consistent with \citet{Naze2019}. Based on the revised
estimate of the radial velocity amplitude, we constrained $M_{2}$ to be below the
Chandrasekhar mass limit over an assumed range of the primary mass and the
inclination angle.

Independently from this, we conducted spectral fitting of the X-ray data using our
NuSTAR and the archived XMM-Newton data with accreting WD models. This is under the
working hypothesis that $\pi$ Aqr hosts a WD. The assumption is validated by the fact
that the fitting yielded high goodness of fit with reasonable best-fit
parameters. Separate models for non-magnetic and magnetic WDs were applied and
constraints on the $M_{2}$ and the inclination angle were obtained. The X-ray results
are closer to the optical result adopting the smaller value of the radial velocity
amplitude. We thus conclude that the secondary of $\pi$ Aqr being a WD is not excluded.

%%%%%%%%%%%%%%%%%%%%%%%%%%%%%%%%%%%%%%%%%%%%%%%%%%%%%%%%%%%%
% Acknowledgments
%%%%%%%%%%%%%%%%%%%%%%%%%%%%%%%%%%%%%%%%%%%%%%%%%%%%%%%%%%%%
\medskip

We appreciate the support from the staff and volunteers to observe with the OAO 188~cm
telescope. We acknowledge Hiroki Harakawa for providing us with the pipeline products of
the HIDES data, Ya\"{e}l Naz\'{e} for interesting discussion, and the anonymous reviewer
for careful review and correction of critical calculation.
This research has made use of the NuSTAR Data Analysis Software (NuSTARDAS) jointly
developed by the ASI Space Science Data Center (SSDC, Italy) and the California
Institute of Technology (Caltech, USA).
This research has made use of the SIMBAD database, operated at CDS, Strasbourg, France
and data from from the European Space Agency (ESA) mission
{\it Gaia} (\url{https://www.cosmos.esa.int/gaia}), processed by the {\it Gaia}
Data Processing and Analysis Consortium (DPAC,
\url{https://www.cosmos.esa.int/web/gaia/dpac/consortium}). Funding for the DPAC
has been provided by national institutions, in particular the institutions
participating in the {\it Gaia} Multilateral Agreement.

\bibliographystyle{aa}
\bibliography{ms}
\end{document}